\documentclass[twocolumn,prc,showpacs,preprintnumbers,
               unsortedaddress,amsmath,amssymb,floatfix]{revtex4}
\usepackage{graphicx}% Include figure files
\usepackage{dcolumn}% Align table columns on decimal point
\usepackage{bm}% bold math
\usepackage{longtable}

\newcommand{\etal}{{\em{et al}}}
\newcommand{\ibid}{{\em{ibid}}}
\newcommand{\tm}{\times}

\begin{document}

%\preprint{JCNP2003-}

\title{Spherical Relativistic Hartree theory in a Woods-Saxon basis}

\author{Shan-Gui Zhou}
 \email{sgzhou@mpi-hd.mpg.de}
 \homepage{http://jcnp.pku.edu.cn/~sgzhou}
 \affiliation{School of Physics, Peking University,
              Beijing 100871, China}
 \affiliation{Max-Planck-Institut f\"ur Kernphysik,
              69029 Heidelberg, Germany\footnote{Present address.}}
 \affiliation{Institute of Theoretical Physics, Chinese Academy of
              Sciences, Beijing 100080, China}
 \affiliation{Center of Theoretical Nuclear Physics, National Laboratory
              of Heavy Ion Accelerator, Lanzhou 730000, China}
\author{Jie Meng}
 \email{mengj@pku.edu.cn}
 \affiliation{School of Physics, Peking University,
              Beijing 100871, China}
 \affiliation{Physikdepartment, Technische Universit\"at M\"unchen,
              85748 Garching, Germany}
 \affiliation{Institute of Theoretical Physics, Chinese Academy of
              Sciences, Beijing 100080, China}
 \affiliation{Center of Theoretical Nuclear Physics, National Laboratory
              of Heavy Ion Accelerator, Lanzhou 730000, China}
\author{P. Ring}
 \email{ring@physik.tu-muenchen.de}
 \affiliation{Physikdepartment, Technische Universit\"at M\"unchen,
              85748 Garching, Germany}

\date{\today}

\begin{abstract}
The Woods-Saxon basis has been suggested to replace the widely
used harmonic oscillator basis for solving the relativistic mean
field (RMF) theory in order to generalize it to study exotic
nuclei. As examples, relativistic Hartree theory is solved for
spherical nuclei in a Woods-Saxon basis obtained by solving either
the Schr\"odinger equation or the Dirac equation (labelled as
SRHSWS and SRHDWS, respectively and SRHWS for both). In SRHDWS,
the negative levels in the Dirac Sea must be properly included.
The basis in SRHDWS could be smaller than that in SRHSWS which
will simplify the deformed problem. The results from SRHWS are
compared in detail with those from solving the spherical
relativistic Hartree theory in the harmonic oscillator basis
(SRHHO) and those in the coordinate space (SRHR). All of these
approaches give identical nuclear properties such as total binding
energies and root mean square radii for stable nuclei. For exotic
nuclei, e.g., $^{72}$Ca, SRHWS satisfactorily reproduces the
neutron density distribution from SRHR, while SRHHO fails. It is
shown that the Woods-Saxon basis can be extended to more
complicated situations for exotic nuclei where both deformation
and pairing have to be taken into account.
\end{abstract}


\maketitle

\section{\label{sec:intro}Introduction}

The existence of an average field in atomic nuclei revealed by
the exceptional role of the nuclear magic numbers provides the
foundation of the nuclear shell model and various mean field
approaches~\cite{Mayer55,Bohr69,Ring80}. This average field is
believed to be approximated most closely by a Woods-Saxon (WS)
potential~\cite{Woods54} either from analyzing the radial
dependence of the nuclear force or by deriving it from a microscopic
two-body force.

Since the eigenfunctions for the WS potential can not be given
analytically, as good approximations for stable nuclei, one often
adopts the harmonic oscillator (HO) potential or the square well,
in particular the former, in shell model calculations for both
spherical~\cite{Mayer55} and deformed nuclei~\cite{Nilsson55}. The
HO eigenfunctions also often serve as a complete basis in solving
equations in both non-relativistic and relativistic mean field
approximations, such as the Skyrme Hartree-Fock (SHF),
Hartree-Fock-Bogoliubov (HFB), relativistic Hartree (RH) and
relativistic Hartree-Bogoliubov (RHB) theories. In these
approaches, solution of corresponding equations is transformed to
a matrix diagnolization problem which can be easily dealt with.

However, due to the incorrect asymptotic property of the HO wave
functions, the expansion in the localized HO basis are not
appropriate for the description of drip line
nuclei~\cite{Dobaczewski96,Meng96,Zhou00} which display many
interesting features because of the extremely weakly bound
property, e.g., the coupling between bound states and the
continuum due to the pairing correlation, large spacial density
distributions, possible modifications of shell structure, \etal.
One must improve the asymptotic behavior of HO wave functions,
e.g., by performing a local scaling
transformation~\cite{Stoitsov98}. However, one does not know the
scaling parameter beforehand thus the predictive power in this
method is lost.

A proper representation to solve the HFB or RHB equations for drip
line nuclei is the coordinate
space~\cite{Dobaczewski84,Meng96,Terasaki96,Poeschl97} where wave
functions are approximated on a spatial lattice and the continuum
is discretized by suitably large box boundary conditions. The HFB
method solved in $r$ space can take fully into account all the
mean-field effects of the coupling to the
continuum~\cite{Dobaczewski96,Meng96,Dobaczewski84,Belyaev87}.
Nevertheless for deformed nuclei, working in $r$ space becomes
much more difficult and numerically very
sophisticated~\cite{Zhou00}. Particularly, it become very time
consuming when the pairing correlation is included. Therefore much
effort is made towards a more efficient solution of HFB or RHB
equations, e.g., using natural orbitals~\cite{Reinhard97} or
working on basis-spline Galerkin lattices~\cite{Oberacker99},
etc..

A reconciler between the HO basis and the $r$ space may be the WS
basis because (i) the WS potential represents the nuclear average
field more suitably than the HO potential and (ii) in principal
there is no localization restrictions in the WS potential.
Although analytical wave functions can not be given for the WS
potential, one may easily find numerical solutions for a spherical
WS potential in the $r$ space by virtue of various effective
methods of solving ordinary differential
equations~\cite{NumericalRecipe}. One can still use a large box
boundary condition to discretize the continuum. These WS wave
functions can thus be used as a complete basis for spherical or
deformed systems and one finally comes back to the familiar matrix
diagnolization problem.

In the present work we restrict the application of this method to
nuclei with spherical symmetry which largely facilitates the
discussion of basic principles and allows presenting illustrations
for the radial dependence of all relevant physical quantities like
density distributions. We combine this approach with the
relativistic Hartree theory~\cite{Serot86} which provides a
framework for describing the nuclear many body problem as a
relativistic system of baryons and mesons and, together with its
extensions with deformation and/or pairing included, have been
successfully applied in calculations of nuclear matter and
properties of finite nuclei throughout the periodic
table~\cite{Reinhard89,Ring96}.

The paper is organized as follows. In Sec.~\ref{sec:formalism}, we
give a brief reminder of the formalism of relativistic Hartree
theory. The numerical details of solving it in the WS basis are
given in Sec.~\ref{sec:numerical}. In Sec.~\ref{sec:results} we
present our results and compare them with those obtained in the HO
basis and in the $r$ space. We also discuss the contribution from
negative levels in the Dirac sea in the same section. Finally, the
work is summarized in Sec.~\ref{sec:summary}.

Throughout the paper, the relativistic Hartree theories solved in
the $r$ space, in the HO basis and in the WS basis are abbreviated
as ``SRHR'', ``SRHHO'' and ``SRHWS'' where the first ``S''
represents ``spherical''. We use ``SWS'' and ``DWS'' to
distinguish the WS basis which is obtained from solving the
Schr\"odinger equation or the Dirac equation with initial WS
potentials, respectively. Thus we have ``SRHSWS'' and ``SRHDWS''
theories.

\section{\label{sec:formalism}Basic formalism of relativistic
Hartree theory}

The starting point of the relativistic Hartree theory is a
Lagrangian density where nucleons are described as Dirac spinors
which interact via the exchange of several mesons ($\sigma$,
$\omega$, and $\rho$) and the
photon~\cite{Serot86,Reinhard89,Ring96},
\begin{eqnarray}
\displaystyle
 {\cal L}
   & = &
     \bar\psi_i \left( i\rlap{/}\partial -M \right) \psi_i
    + \frac{1}{2} \partial_\mu \sigma \partial^\mu \sigma
    - U(\sigma)
    - g_{\sigma} \bar\psi_i \sigma \psi_i
   \nonumber \\
   &   & \mbox{}
    - \frac{1}{4} \Omega_{\mu\nu} \Omega^{\mu\nu}
    + \frac{1}{2} m_\omega^2 \omega_\mu \omega^\mu
    - g_{\omega} \bar\psi_i \rlap{/}\omega \psi_i
   \nonumber \\
   &   & \mbox{}
    - \frac{1}{4} \vec{R}_{\mu\nu} \vec{R}^{\mu\nu}
    + \frac{1}{2} m_{\rho}^{2} \vec{\rho}_\mu \vec{\rho}^\mu
    - g_{\rho} \bar\psi_i \rlap{/} \vec{\rho} \vec{\tau} \psi_i
   \nonumber \\
   &   &\mbox{}
    - \frac{1}{4} F_{\mu\nu} F^{\mu\nu}
    - e \bar\psi_i \frac{1-\tau_3}{2}\rlap{/}A \psi_i ,
\label{eq:Lagrangian}
\end{eqnarray}
with the summation convention used and the summation over $i$ runs
over all nucleons, $\rlap{/} x \equiv \gamma^\mu x_\mu =
\gamma_\mu x^\mu$, $M$ the nucleon mass, and $m_\sigma$,
$g_\sigma$, $m_\omega$, $g_\omega$, $m_\rho$, $g_\rho$ masses and
coupling constants of the respective mesons. The nonlinear
self-coupling for the scalar mesons is given by~\cite{Boguta77}
\begin{equation}
   U(\sigma) = \dfrac{1}{2} m^2_\sigma \sigma^2
              +\dfrac{g_2}{3}\sigma^3 + \dfrac{g_3}{4}\sigma^4 ,
\end{equation}
and field tensors for the vector mesons and the photon fields are
defined as
\begin{eqnarray}
 \left\{
  \begin{array}{rcl}
   \Omega_{\mu\nu}  & = & \partial_\mu\omega_\nu
                         -\partial_\nu\omega_\mu, \\
   \vec{R}_{\mu\nu} & = & \partial_\mu\vec{\rho}_\nu
                         -\partial_\nu\vec{\rho}_\mu
                         -g_{\rho} (\vec{\rho}_\mu
                                    \tm \vec{\rho}_\nu ), \\
   F_{\mu\nu}       & = & \partial_\mu {A}_\nu
                         - \partial_\nu {A}_\mu.
  \end{array}
 \right.
 \label{eq:tensors}
\end{eqnarray}

The classical variation principle gives the equations of motion
for the nucleons, mesons and the photon. As in many applications,
we study the ground state properties of nuclei with time reversal
symmetry, thus the nucleon spinors are the eigenvectors of the
stationary Dirac equation
\begin{equation}
  \left[ \bm{\alpha} \cdot \bm{p} + V(\bm{r}) + \beta (M + S(\bm{r}))
  \right] \psi_i(\bm{r}) = \epsilon_i \psi_i(\bm{r}) ,
\label{eq:Dirac0}
\end{equation}
and equations of motion for the mesons and the photon are
\begin{eqnarray}
 \left\{
   \begin{array}{rcl}
    \left( -\Delta + \partial_\sigma U(\sigma) \right )\sigma(\bm{r})
      & = & -g_\sigma \rho_s(\bm{r}) , \\
    \left( -\Delta + m_\omega^2 \right )             \omega^0(\bm{r})
      & = &  g_\omega \rho_v(\bm{r}) , \\
    \left( -\Delta + m_\rho^2 \right)                  \rho^0(\bm{r})
      & = &  g_\rho   \rho_3(\bm{r}) , \\
    -\Delta                                               A^0(\bm{r})
      & = &  e        \rho_p(\bm{r}) ,
   \end{array}
 \right.
 \label{eq:mesonmotion}
\end{eqnarray}
where $\omega^0$ and $A^0$ are time-like components of the vector
$\omega$ and the photon fields and $\rho^0$ the 3-component of the
time-like component of the iso-vector vector $\rho$ meson.
Equations~(\ref{eq:Dirac0}) and (\ref{eq:mesonmotion}) are coupled
by the vector and scalar potentials
\begin{eqnarray}
 \left\{
   \begin{array}{lll}
     V(\bm{r}) & = & g_\omega \omega^0(\bm{r})
                    +g_\rho \tau_3 \rho^0(\bm{r})
                    +e \dfrac{1-\tau_3}{2} A^0(\bm{r}) , \\
     S(\bm{r}) & = & g_\sigma \sigma(\bm{r}), \\
   \end{array}
 \right.
 \label{eq:vaspot}
\end{eqnarray}
and various densities
\begin{eqnarray}
 \left\{
  \begin{array}{rcl}
   \rho_s(\bm{r})
   & = &
    \sum_{i=1}^A \bar\psi_i(\bm{r}) \psi_i(\bm{r}) ,\\
   \rho_v(\bm{r})
   & = &
    \sum_{i=1}^A \psi_i^\dagger(\bm{r}) \psi_i(\bm{r}) ,\\
   \rho_3(\bm{r})
   & = &
    \sum_{i=1}^A \psi_i^\dagger(\bm{r}) \tau_3 \psi_i(\bm{r}) ,\\
   \rho_c(\bm{r})
   & = &
    \sum_{i=1}^A \psi_i^\dagger(\bm{r})
                 \dfrac{1-\tau_3}{2}\psi_i(\bm{r}) .
  \end{array}
 \right.
 \label{eq:mesonsource}
\end{eqnarray}

For spherical nuclei, meson fields and densities depend only on
the radial coordinate $r$. The spinor is characterized by the
angular momentum quantum numbers ($l$,$j$), $m$, the parity, the
isospin $t = \pm 1/2$ (``+'' for neutrons and ``$-$'' for
protons), and the radial quantum number $\alpha$. The Dirac spinor
has the form
\begin{equation}
 \psi_{\alpha\kappa m}(\bm{r},s,t) =
   \left(
     \begin{array}{c}
       i \dfrac{G_\alpha^{\kappa}(r)}{r} Y^l _{jm} (\theta,\phi)
       \\
       - \dfrac{F_\alpha^{\kappa}(r)}{r} Y^{\tilde l}_{jm}(\theta,\phi)
     \end{array}
   \right) \chi_{t_\alpha}(t),
   \ \ j = l\pm\frac{1}{2},
 \label{eq:SRHspinor}
\end{equation}
with $G_\alpha^{\kappa}(r) / r$ and $F_\alpha^{\kappa}(r) / r$ the
radial wave functions for the upper and lower components and $Y^l
_{jm}(\theta,\phi)$ the spin spherical harmonics where $\kappa =
(-1)^{j+l+1/2} (j+1/2)$ and $\tilde l = l + (-1)^{j+l-1/2}$. The
value of $\kappa$ of the {\em upper} component is used to label a
state both for normal levels in the Fermi sea and for negative
ones in the Dirac sea. States with the same $\kappa$ form a
``block''. The radial equation of the Dirac spinor, Eq.
(\ref{eq:Dirac0}), is reduced as
%\begin{widetext}
\begin{equation}
 \left\{
   \begin{array}{lll}
    \epsilon_\alpha G_\alpha^{\kappa} & = &
     \left( -\dfrac{\partial}{\partial r} + \dfrac{\kappa}{r}
     \right) F_\alpha^{\kappa}
     + \left( M + S(r) + V(r) \right) G_\alpha^{\kappa} ,
    \\
    \epsilon_\alpha F_\alpha^{\kappa} & = &
     \left( +\dfrac{\partial}{\partial r} + \dfrac{\kappa}{r}
     \right) G_\alpha^{\kappa}
     - \left( M + S(r) - V(r) \right) F_\alpha^{\kappa} .
   \end{array}
 \right.
 \label{eq:SRHDirac}
\end{equation}
%\end{widetext}
The meson field equations become simply radial Laplace equations
of the form
\begin{equation}
 \left( - \frac{\partial^2}{\partial r^2}
        - \frac{2}{r}\frac{\partial}{\partial r} + m_{\phi}^2
 \right) \phi(r)
 = s_{\phi}(r).
 \label{eq:SRHmesonmotion}
\end{equation}
$m_{\phi}$ are the meson masses for $\phi = \sigma, \omega,\rho$
and zero for the photon. The source terms are
\begin{eqnarray}
 s_{\phi}(r) =
  \left\{
    \begin{array}{ll}
      - g_\sigma\rho_s(r) - g_2 \sigma^2(r)  - g_3 \sigma^3(r),
     & \text{for}\ \sigma, \\
        g_\omega \rho_v(r),
     & \text{for}\ \omega, \\
        g_{\rho} \rho_3(r),
     & \text{for}\ \rho, \\
        e \rho_c(r),
     & \text{for}\ A, \\
    \end{array}
  \right.
 \label{eq:sources}
\end{eqnarray}
with
\begin{eqnarray}
 \left\{
  \begin{array}{lll}
   4\pi r^2 \rho_s(r) & = & \sum_{i=1}^A (|G_i(r)|^2 - |F_i(r)|^2), \\
   4\pi r^2 \rho_v(r) & = & \sum_{i=1}^A (|G_i(r)|^2 + |F_i(r)|^2), \\
   4\pi r^2 \rho_3(r) & = & \sum_{i=1}^A 2t_i
                                         (|G_i(r)|^2 + |F_i(r)|^2), \\
   4\pi r^2 \rho_c(r) & = & \sum_{i=1}^A \left(\frac{1}{2}-t_i\right)
                                         (|G_i(r)|^2 + |F_i(r)|^2). \\
  \end{array}
 \right.
 \label{eq:mesonsourceS}
\end{eqnarray}

\begin{table}
\caption{\label{tab:srhsws-dr}Dependence of the average single
particle energy, root mean square (rms) radius, and $\langle
r^4\rangle^{1/4}$ on the mesh size $\Delta r$ for the SRHSWS
theory. The meson and Coulomb fields are obtained from SRHR
calculations with the parameter set NL3, $\Delta r = 0.05$ fm,
$R_\text{max}$ = 30 fm for $^{16}$O and 35 fm for $^{208}$Pb. In
SRHSWS calculations, the parameter set NL3 is used. For $^{16}$O,
$R_\text{max} = 4r_0A^{1/3} = 12.8$ fm, $E_\text{cut} = 300$ MeV.
For $^{208}$Pb, $R_\text{max} = 3r_0A^{1/3} = 22.6$ fm,
$E_\text{cut} = 200$ MeV. The first row gives results from solving
the Dirac equation in the coordinate space.}
\begin{ruledtabular}
\begin{tabular}{c|ccc|ccc}
 $\Delta r$ &
 $-E_\text{sp}/A$ & $\langle r^2\rangle^{1/2}$ &
 $\langle r^4\rangle^{1/4}$ &
 $-E_\text{sp}/A$ & $\langle r^2\rangle^{1/2}$ &
 $\langle r^4\rangle^{1/4}$ \\
\hline
 & \multicolumn{3}{c|}{$^{16}$O} & \multicolumn{3}{c}{$^{208}$Pb} \\
 ---  & 23.0375 & 2.5945 & 2.8900 & 23.3348 & 5.6315 & 5.9883 \\
 0.05 & 23.0375 & 2.5945 & 2.8900 & 23.3348 & 5.6315 & 5.9883 \\
 0.10 & 23.0376 & 2.5945 & 2.8899 & 23.3348 & 5.6315 & 5.9883 \\
 0.20 & 23.0385 & 2.5941 & 2.8896 & 23.3347 & 5.6315 & 5.9883 \\
 0.30 & 23.0430 & 2.5930 & 2.8889 & 23.3343 & 5.6314 & 5.9883 \\
 0.40 & 23.0420 & 2.5923 & 2.8885 & 23.3334 & 5.6315 & 5.9885 \\
 0.50 & 22.9887 & 2.5949 & 2.8936 & 23.3284 & 5.6319 & 5.9890 \\
\end{tabular}
\end{ruledtabular}
\end{table}

The above coupled equations have been solved in the $r$
space~\cite{Horowitz81} and in the HO basis~\cite{Gambhir90} with
the no sea and the mean field approximation. Here we depict
briefly the procedure of solving these coupled equations. With a
set of estimated meson and photon fields, the scalar and vector
potentials are calculated and the radial Dirac equation is solved.
Thus obtained nucleon wave functions are used to calculate the
source term of each radial Laplace equation for mesons and the
photon. New meson and photon fields are calculated from solving
these Laplace equations. This procedure is iterated until a
demanded accuracy is achieved. Laplace equations are usually
solved using the Green's function
method~\cite{Horowitz81,Gambhir90} though in Ref.~\cite{Gambhir90}
Laplace equations for mesons are solved in the HO basis. SRHR,
SRHHO and SRHWS differ from each other mainly in how to solve the
Dirac equation. In the following the numerical solution of the
Dirac equation in the WS basis will be presented.

\section{\label{sec:numerical} Solving the Dirac equation in a
Woods-Saxon basis and numerical details}

\subsection{\label{subsec:srhsws}Woods-Saxon basis from solving
a Schr\"odinger equation (the SWS basis)}

For the Schr\"odinger equation with a spherical Woods-Saxon
potential
\begin{equation}
 V_\text{WS}(r)
  = \left\{
     \begin{array}{ll}
      \dfrac{V_0}{1+e^{(r-R_0)/a_0}},\ \ & r <    R_\text{max}, \\
      \infty,                            & r \geq R_\text{max}, \\
     \end{array}
    \right.
\end{equation}
where $R_\text{max}$ is introduced for practical reasons to define
the box boundary, the eigenfunction can be written as
$\phi_{nlm_l}(\bm{r}) = R_{nl}(r) Y_{lm_l}(\theta,\phi)$. Its
radial Schr\"odinger equation is derived as
\begin{eqnarray}
 & &
 \left[ -\dfrac{1}{2M}
  \left( \dfrac{1}{r^2} \dfrac{\partial}{\partial r}
         r^2 \dfrac{\partial}{\partial r} - \dfrac{l(l+1)}{r^2}
  \right)
  + V_\text{WS}(r)
 \right] R_{nl}(r)
 \nonumber \\
 & &
 =\ E_{nl} R_{nl}(r) .
 \label{eq:schws}
\end{eqnarray}

\begin{table}
\caption{\label{tab:srhsws-dn}Dependence of the average single
particle energy, rms radius, and $\langle r^4\rangle^{1/4}$ on the
difference $\Delta n=\tilde n_\text{max}-n_\text{max}$ for the
SRHSWS theory. The meson and Coulomb fields are obtained from SRHR
calculations with the parameter set NL3, $\Delta r = 0.05$ fm,
$R_\text{max}$ = 30 fm for $^{16}$O and 35 fm for $^{208}$Pb. In
SRHSWS calculations, the parameter set NL3 is used. For $^{16}$O,
$R_\text{max} = 4r_0A^{1/3} = 12.8$ fm. For $^{208}$Pb,
$R_\text{max} = 3r_0A^{1/3} = 22.6$ fm.}
\begin{ruledtabular}
\begin{tabular}{c|ccc|ccc}
 $\Delta n$ &
 $-E_\text{sp}/A$ & $\langle r^2\rangle^{1/2}$ &
 $\langle r^4\rangle^{1/4}$ &
 $-E_\text{sp}/A$ & $\langle r^2\rangle^{1/2}$ &
 $\langle r^4\rangle^{1/4}$ \\
\hline
 &
 \multicolumn{3}{c}{$^{16}$O: $E_\text{cut} = 100$ MeV} &
 \multicolumn{3}{c}{$^{208}$Pb: $E_\text{cut} = 100$ MeV} \\
  1 & 23.0382 & 2.5947 & 2.8920 & 23.3344 & 5.6315 & 5.9884 \\
  3 & 23.0326 & 2.5949 & 2.8913 & 23.3341 & 5.6315 & 5.9884 \\
  5 & 23.0298 & 2.5952 & 2.8915 & 23.3340 & 5.6315 & 5.9884 \\
  7 & 23.0290 & 2.5953 & 2.8915 & 23.3340 & 5.6315 & 5.9884 \\
  9 & 23.0289 & 2.5953 & 2.8915 & 23.3340 & 5.6315 & 5.9884 \\
\hline
 &
 \multicolumn{3}{c}{$^{16}$O: $E_\text{cut} = 300$ MeV} &
 \multicolumn{3}{c}{$^{208}$Pb: $E_\text{cut} = 200$ MeV} \\
  1 & 23.0375 & 2.5945 & 2.8999 & 23.3348 & 5.6315 & 5.9883 \\
  3 & 23.0375 & 2.5945 & 2.8999 & 23.3348 & 5.6315 & 5.9883 \\
  5 & 23.0375 & 2.5945 & 2.8999 & 23.3347 & 5.6315 & 5.9883 \\
  7 & 23.0375 & 2.5945 & 2.8999 & 23.3347 & 5.6315 & 5.9883 \\
  9 & 23.0375 & 2.5945 & 2.8999 & 23.3347 & 5.6315 & 5.9883 \\
\end{tabular}
\end{ruledtabular}
\end{table}

Equation~(\ref{eq:schws}) is solved on a discretized radial mesh
with a mesh size $\Delta r$. $R_\text{max}$ ($\Delta r$) should be
chosen larger (smaller) enough to make sure that the final results
do not depend on it. The radial wave functions thus obtained form
a complete basis,
\begin{equation}
 \left\{ R_{nl}(r); n=0, 1, \cdots; l = 0,1,\cdots,n \right\},
\end{equation}
in terms of which the radial part of the upper and the lower
components of the Dirac spinor in Eq.~(\ref{eq:SRHDirac}) are
expanded respectively as
\begin{equation}
 \left\{
  \begin{array}{l}
   \displaystyle
   G_\alpha^\kappa(r) = -i \sum_{n=0}^{n_\text{max}}
                             g_{\alpha n} r R_{nl}(r), \\
   \displaystyle
   F_\alpha^\kappa(r) = -i \sum_{\tilde n=0}^{\tilde n_\text{max}}
                             f_{\alpha \tilde n}
                             r R_{\tilde n\tilde l}(r). \\
  \end{array}
 \right. \label{eq:DiracSWS}
\end{equation}
The radial Dirac equation, Eq.~(\ref{eq:SRHDirac}), is transformed
into the WS basis as
\begin{equation}
 \left(
  \begin{array}{cc}
   {\cal A}_{mn}        & {\cal B}_{m\tilde n}        \\
   {\cal C}_{\tilde mn} & {\cal D}_{\tilde m\tilde n} \\
  \end{array}
 \right)
 \left(
  \begin{array}{c}
   g_{\alpha n} \\
   f_{\alpha \tilde n} \\
  \end{array}
 \right)
 = \epsilon_\alpha
 \left(
  \begin{array}{c}
   g_{\alpha n} \\
   f_{\alpha \tilde n} \\
  \end{array}
 \right),
\end{equation}
where the matrix elements are calculated as follows
\begin{equation}
 \left\{
  \begin{array}{l}
   \displaystyle
   {\cal A}_{mn}
     = \int_0^{R_\text{max}} r^2 dr
        R_{ml}(r)
        \left( V(r)+S(r)+M \right)
        R_{nl}(r),
   \\
   \displaystyle
   {\cal B}_{m\tilde n}
     = \int_0^{R_\text{max}} r^2 dr
        R_{ml}(r)
        \left( +\frac{\partial}{\partial r} - \frac{\kappa_\alpha-1}{r}
        \right)
        R_{\tilde n\tilde l}(r),
   \\
   \displaystyle
   {\cal C}_{\tilde mn}
     = \int_0^{R_\text{max}} r^2 dr
        R_{\tilde m\tilde l}(r)
        \left( -\frac{\partial}{\partial r} - \frac{\kappa_\alpha+1}{r}
        \right)
        R_{nl}(r),
   \\
   \displaystyle
   {\cal D}_{\tilde m\tilde n}
     = \int_0^{R_\text{max}} r^2 dr
        R_{\tilde m\tilde l}(r)
        \left( V(r)-S(r)-M \right)
        R_{\tilde n\tilde l}(r).
   \\
  \end{array}
 \right.
 \label{eq:DiracMatrix}
\end{equation}

In practical calculations, an energy cutoff $E_\text{cut}$
(relative to the nucleon mass $M$) is used to determine the cutoff
of the radial quantum number $n_\text{max}$ for each block. In the
expansion of the corresponding lower component, we take $\tilde
n_\text{max} = n_\text{max} + \Delta n$ with $\Delta n\ge 1$ in
order to avoid spurious states~\cite{Gambhir90}.

\begin{figure}
\includegraphics[width=8.5cm]{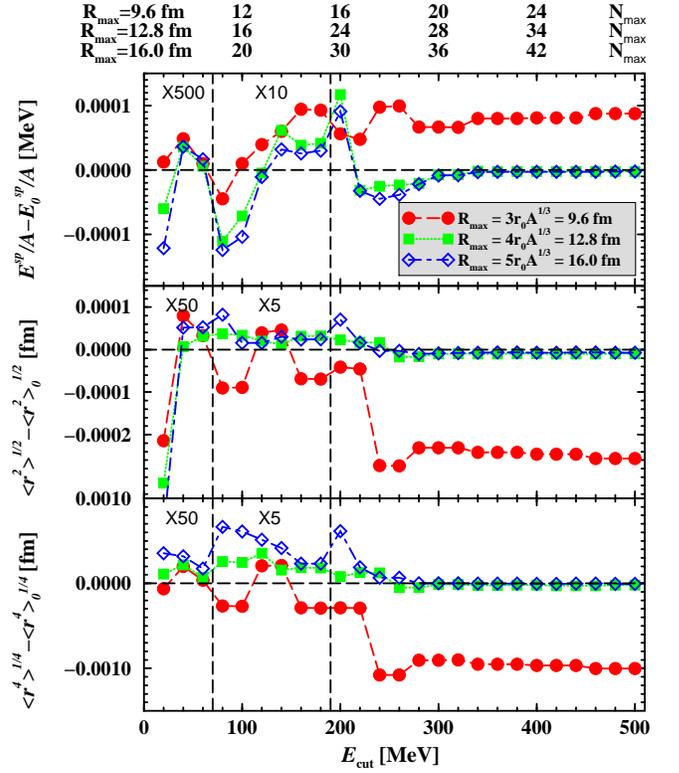}
\caption{\label{fig:sconvs.o16} Deviations of the average single
particle energy $E^\text{sp}/A$ (the upper panel), rms radius
$\langle r\rangle^{1/2}$ (the middle panel), and $\langle
r^4\rangle^{1/4}$ (the lower panel) of $^{16}$O from the standard
results versus the cutoff energy $E_\text{cut}$ with different box
size $R_\text{max}$ for the SRHSWS theory. The meson and Coulomb
fields are obtained from SRHR calculations with the parameter set
NL3, $\Delta r = 0.05$ fm, $R_\text{max}$ = 30 fm. In SRHSWS
calculations, the parameter set NL3 is used. For $E_\text{cut}$ =
100, 200, 300, 400 MeV, the approximate maximum principal quantum
number in each basis, $N_\text{max} = 2n_\text{max}+l$, is given
on the top of the plot.}
\end{figure}

\begin{figure}
\includegraphics[width=8.5cm]{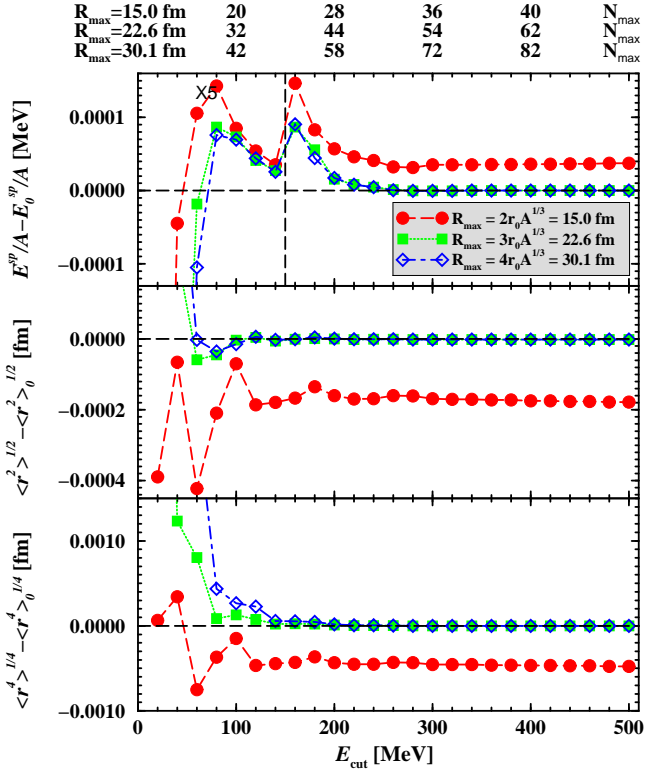}
\caption{\label{fig:sconvs.pb208} Deviations of the average single
particle energy $E^\text{sp}/A$ (the upper panel), rms radius
$\langle r\rangle^{1/2}$ (the middle panel), and $\langle
r^4\rangle^{1/4}$ (the lower panel) of $^{208}$Pb from the
standard results versus the cutoff energy $E_\text{cut}$ with
different box size $R_\text{max}$ for the SRHSWS theory. The meson
and Coulomb fields are obtained from SRHR calculations with the
parameter set NL3, $\Delta r = 0.05$ fm, $R_\text{max}$ = 35 fm.
In SRHSWS calculations, the parameter set NL3 is used. For
$E_\text{cut}$ = 100, 200, 300, 400 MeV, the approximate maximum
principal quantum number in each basis, $N_\text{max} =
2n_\text{max}+l$, is given on the top of the plot.}
\end{figure}

The following Woods-Saxon parameters have been used according to
Ref.~\cite{Heyde99}
\begin{equation}
 \left\{
  \begin{array}{l}
   V_0 = (-51 \pm 33(N-Z)/A)
    \text{ MeV},\\
   R_0 = 1.27 A^{1/3} \text{ fm},\ a_0 = 0.67 \text{ fm}, \\
  \end{array}
 \right.
\end{equation}
where `+' is for the neutron and `$-$' for the proton. As
expected, the dependence of final results on the initial WS
potential is almost negligible. For example, a variation of $V_0$
by 50\% gives differences in total binding energies by less than
0.1\% and charge radii by less than 0.5\% for $^{16}$O, $^{48}$Ca
and $^{208}$Pb. Such situation is also checked to be true for the
other two parameters in the WS potential, $R_0$ and $a_0$.

\subsection{\label{subsec:srhdws}Woods-Saxon basis from solving
a Dirac equation (the DWS basis)}

The radial Dirac equation, Eq.~(\ref{eq:SRHDirac}), may be solved
in the $r$ space~\cite{Horowitz81} with Woods-Saxon-like
potentials for $V_0(r) \pm S_0(r)$~\cite{Koepf91} within a
spherical box of the size $R_\text{max}$, together with the
spherical spinor which gives a complete WS basis
%\begin{widetext}
\begin{equation}
 \left\{
  \left[ \epsilon_{n\kappa m}^0, \psi_{n\kappa m}^0(\bm{r},s,t) \right];
   \epsilon_{n\kappa m}^0 \gtrless 0
 \right\},
\end{equation}
%\end{widetext}
with $n=0,1,\cdots$, $\kappa=\pm1,\pm2,\cdots$, and $m =
-j_\kappa,\cdots,j_\kappa$. $\psi_{n\kappa m}^0(\bm{r},s,t)$ takes
the form of Eq.~(\ref{eq:SRHspinor}). We note that states both in
the Fermi sea and in the Dirac sea should be included in the basis
for the completeness. The nucleon wave function,
Eq.~(\ref{eq:SRHspinor}), can be expanded in terms of this set of
basis as
\begin{equation}
 \psi_{\alpha\kappa m}(\bm{r},s,t)
  = \sum_{n=0}^{n_\text{max}}
     c_{\alpha n} \psi_{n\kappa m}^0(\bm{r},s,t),
 \label{eq:DSexpansion}
\end{equation}
where $n_\text{max} = n_\text{max}^+ + n_\text{max}^- + 1$ and the
summation is over normal levels in the Fermi sea for $0 \le n \le
n_\text{max}^+$ and over negative levels in the Dirac sea for
$n_\text{max}^+ + 1 \le n \le n_\text{max}$. The negative states
is obtained with the same method as the positive
ones~\cite{Horowitz81}. In this WS basis, the Dirac equation,
Eq.~(\ref{eq:Dirac0}), turns out to be
\begin{equation}
 c_{\alpha m} \epsilon_m^0 + \sum_{n=0}^{n_\text{max}}
                              c_{\alpha n} H'_{mn}
 = \epsilon_\alpha c_{\alpha m},\
 m = 1, \cdots, n_\text{max},
\label{eq:matrix}
\end{equation}
with
%\begin{widetext}
\begin{eqnarray}
 H'_{mn}
 & = &
  \left\langle \psi_m^0(\bm{r}) \right|
   \left[ \Delta V(\bm{r})+\beta\Delta S(\bm{r})\right]
  \left| \psi_n^0(\bm{r}) \right\rangle
 \nonumber \\
 & = &
  \int_0^{R_\text{max}} dr
    G_m^0(r) \left[ \Delta V(r)+\Delta S(r) \right] G_n^0(r)
 \nonumber \\
 & + &
  \int_0^{R_\text{max}} dr
    F_m^0(r) \left[ \Delta V(r)-\Delta S(r) \right] F_n^0(r)
 ,
 \label{eq:matrixelement}
\end{eqnarray}
%\end{widetext}
where $\Delta V(\bm{r}) = V(\bm{r})-V_0(\bm{r})$ and $\Delta
S(\bm{r}) = S(\bm{r})-S_0(\bm{r})$. The angular, spin, and isospin
quantum numbers are omitted for brevity.

\begin{figure}
\includegraphics[width=8.5cm]{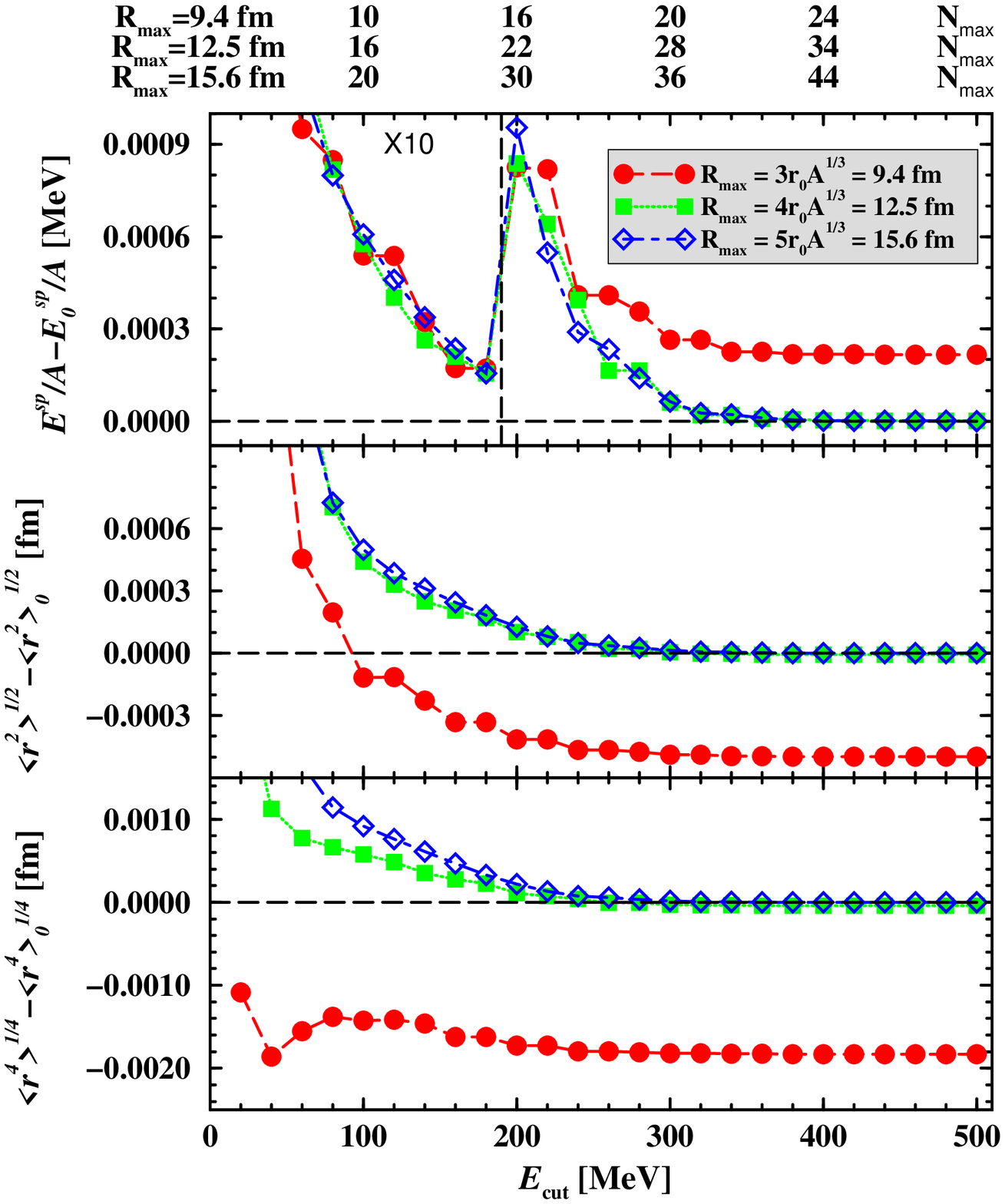}
\caption{\label{fig:sconvd.o16} Deviations of the average single
particle energy $E^\text{sp}/A$ (the upper panel), rms radius
$\langle r\rangle^{1/2}$ (the middle panel), and $\langle
r^4\rangle^{1/4}$ (the lower panel) of $^{16}$O from the standard
results versus the cutoff energy $E_\text{cut}$ with different box
size $R_\text{max}$ for the SRHDWS theory. The meson and Coulomb
fields are obtained from SRHR calculations with the parameter set
NL3, $\Delta r = 0.05$ fm, $R_\text{max}$ = 30 fm. In SRHDWS
calculations, the parameter set NL3 is used. For $E_\text{cut}$ =
100, 200, 300, 400 MeV, the approximate maximum principal quantum
number in each basis, $N_\text{max} = 2n_\text{max}+l$, is given
on the top of the plot.}
\end{figure}

\begin{figure}
\includegraphics[width=8.5cm]{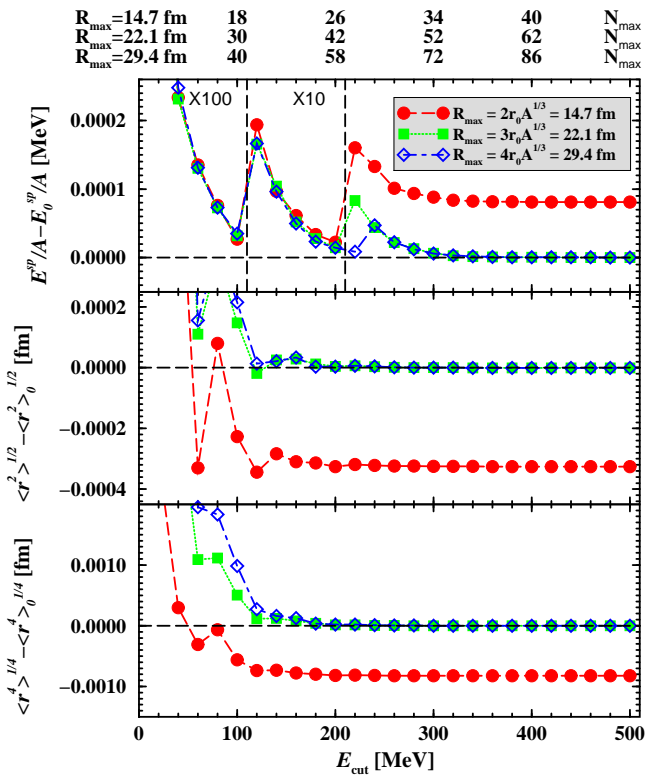}
\caption{\label{fig:sconvd.pb208} Deviations of the average single
particle energy $E^\text{sp}/A$ (the upper panel), rms radius
$\langle r\rangle^{1/2}$ (the middle panel), and $\langle
r^4\rangle^{1/4}$ (the lower panel) of $^{208}$Pb from the
standard results versus the cutoff energy $E_\text{cut}$ with
different box size $R_\text{max}$ for the SRHDWS theory. The meson
and Coulomb fields are obtained from SRHR calculations with the
parameter set NL3, $\Delta r = 0.05$ fm, $R_\text{max}$ = 35 fm.
In SRHDWS calculations, the parameter set NL3 is used. For
$E_\text{cut}$ = 100, 200, 300, 400 MeV, the approximate maximum
principal quantum number in each basis, $N_\text{max} =
2n_\text{max}+l$, is given on the top of the plot.}
\end{figure}

It should be mentioned that Eq.~(\ref{eq:SRHDirac}) can be solved
directly in the $r$ space with the same method of generating the
DWS basis. It is our aim to test the validity of an efficient
solution not only for the spherical RH model but also for its
extension to include the deformation and/or the pairing
correlation. In fact, if only the SRH theory is concerned, this
procedure is just a replacement of the direct solution in the $r$
space by a diagnolization of a matrix with some complication
introduced by the fact that contributions from states in the Dirac
sea must be included.

An energy cutoff $E_\text{cut}$ (relative to the nucleon mass $M$)
and the cutoff of radial quantum numbers $n_\text{max}^+$ are
applied to normal levels alternatively according to practical
convenience. For the initial Woods-Saxon potentials $V_0(r)\pm
S_0(r)$, we follow Ref.~\cite{Koepf91}.

\subsection{Comparison with the $r$-space method}

In order to check the validity of solving the Dirac equation in
the WS basis and to provide numerical experiences for future
applications, we compare results of $^{16}$O and $^{208}$Pb from
solving the Dirac equation in the WS basis and those from solving
the same equation in the coordinate space. The latter is the most
accurate method of solving the Dirac equation for realistic nuclei
up to now thus is used as a standard here. The scalar and vector
potentials in the Dirac equation are provided by very accurate
SRHR calculations with the parameter set NL3 for the Lagrangian,
the mesh size $\Delta r$ = 0.05 fm, the box size $R_\text{max}$ =
30 fm for $^{16}$O and $R_\text{max}$ = 35 fm for $^{208}$Pb. Then
with thus obtained $S(r)$ and $V(r)$, the Dirac equation is solved
in the coordinate space and in the WS basis with also the
parameter set NL3.

To compare results only from solving the Dirac equation avoids
errors from other numerical procedures, e.g., the error from the
iteration and that from solving the Laplace equations. For the
same reason, what we compare between these two methods is not the
binding energy which contains the contribution from mesons but the
average single particle energy $E^\text{sp}/A$. $E^\text{sp} =
\sum_i \epsilon_i$ where $\epsilon_i$ is the single particle
energy and the summation over all occupied states for both
neutrons and protons. We also compare the rms radius $\langle
r^2\rangle^{1/2}$ and $\langle r^4\rangle^{1/4}$. The radius
$\langle r^4\rangle^{1/4}$ reflects the nucleon densities in the
tail more than the rms radius.

Table~\ref{tab:srhsws-dr} presents the dependence of results of
the Dirac equation in the SWS basis on the mesh size $\Delta r$.
With $\Delta r$ decreasing, results in the SWS basis approach the
standard results, i.e., those in the $r$ space. $\Delta r$ = 0.1
fm gives results accurate enough. The dependence on the box size
$R_\text{max}$ and on the basis size determined by $E_\text{cut}$
are investigated and shown in Figs.~\ref{fig:sconvs.o16} and
\ref{fig:sconvs.pb208} where the deviations of the average single
particle energy $E^\text{sp}/A$, the rms radius $\langle
r^2\rangle^{1/2}$ and $\langle r^4\rangle^{1/4}$ from the
standards are plotted versus $E_\text{cut}$ for different
$R_\text{max}$. If $R_\text{max}$ is not large enough, it is
difficult to approach the standard results. For example, when
$R_\text{max} = 3r_0A^{1/3} = 9.4$ fm and $E_\text{cut}$ = 300 MeV
for $^{16}$O (correspondingly, $N_\text{max}\sim 20$), the results
seem converge, but the discrepancy of the average single particle
energy from the standard one remains 0.1 keV. So one must use a
large enough box with the size $R_\text{max}$ around $4r_0A^{1/3}$
for light nuclei and $3r_0A^{1/3}$ for heavy ones. It is
interesting that the convergence of the results does not depend on
$N_\text{max}$ but only on $E_\text{cut}$. For $^{16}$O
($^{208}$Pb), the results converge to the standard ones at $\sim$
300 (400) MeV. From Figs.~\ref{fig:sconvs.o16} and
\ref{fig:sconvs.pb208}, we find the radius $\langle
r^4\rangle^{1/4}$ also converges very well which implies that
nucleon densities could be calculated accurately even for large
$r$.

We have made similar investigations for results in the DWS basis
and similar conclusions are made. For instances, the deviations of
the average single particle energy $E^\text{sp}/A$, the rms radius
$\langle r^2\rangle^{1/2}$ and $\langle r^4\rangle^{1/4}$ from the
standards are plotted versus $E_\text{cut}$ for different
$R_\text{max}$ in Figs.~\ref{fig:sconvd.o16} and
\ref{fig:sconvd.pb208}.

\begin{table}
\caption{\label{tab:convnn}Dependence of the single particle
energy, rms radius, and $\langle r^4\rangle^{1/4}$ on the maximum
principal quantum number, $N_\text{max}^- = 2n_\text{max}^- + l$,
for the SRHDWS theory. The meson and Coulomb fields are obtained
from SRHR calculations with the parameter set NL3, $\Delta r =
0.05$ fm, $R_\text{max}$ = 30 fm for $^{16}$O and 35 fm for
$^{208}$Pb. In SRHDWS calculations, the parameter set NL3 is used.
For $^{16}$O, $R_\text{max} = 4r_0A^{1/3} = 12.5$ fm,
$E_\text{cut} = 300$ MeV for positive states. For $^{208}$Pb,
$R_\text{max} = 3r_0A^{1/3} = 22.1$ fm, $E_\text{cut} = 200$ MeV
for positive states.}
\begin{ruledtabular}
\begin{tabular}{c|ccc|ccc}
 $\tilde N_\text{max}$ &
 $-E_\text{sp}/A$ & $\langle r^2\rangle^{1/2}$ &
 $\langle r^4\rangle^{1/4}$ &
 $-E_\text{sp}/A$ & $\langle r^2\rangle^{1/2}$ &
 $\langle r^4\rangle^{1/4}$ \\
\hline
     & \multicolumn{3}{c}{$^{16}$O}   & \multicolumn{3}{c}{$^{208}$Pb} \\
  no & 23.1129 &   2.5912   &  2.8859 & 23.3331 &   5.6314  &   5.9889 \\
  0  & 23.1077 &   2.5916   &  2.8861 & 23.3329 &   5.6314  &   5.9889 \\
  2  & 23.0762 &   2.5939   &  2.8889 & 23.3316 &   5.6315  &   5.9890 \\
  4  & 23.0617 &   2.5942   &  2.8893 & 23.3304 &   5.6317  &   5.9892 \\
  6  & 23.0439 &   2.5946   &  2.8898 & 23.3299 &   5.6318  &   5.9893 \\
  8  & 23.0385 &   2.5946   &  2.8899 & 23.3294 &   5.6319  &   5.9893 \\
 10  & 23.0376 &   2.5946   &  2.8899 & 23.3292 &   5.6319  &   5.9894 \\
 12  & 23.0375 &   2.5946   &  2.8899 & 23.3291 &   5.6319  &   5.9899 \\
 14  & 23.0375 &   2.5946   &  2.8899 & 23.3290 &   5.6319  &   5.9894 \\
 16  & 23.0375 &   2.5946   &  2.8899 & 23.3290 &   5.6319  &   5.9894 \\
 18  & 23.0375 &   2.5946   &  2.8899 & 23.3289 &   5.6319  &   5.9893 \\
 20  & 23.0375 &   2.5946   &  2.8899 & 23.3288 &   5.6319  &   5.9893 \\
 22  & 23.0375 &   2.5946   &  2.8899 & 23.3287 &   5.6319  &   5.9893 \\
 30  & 23.0375 &   2.5946   &  2.8899 & 23.3287 &   5.6319  &   5.9893 \\
\end{tabular}
\end{ruledtabular}
\end{table}

In the expansion of the nucleon wave function,
Eq.~(\ref{eq:DSexpansion}), one has to take into account not only
the levels in the Fermi sea but also those in the Dirac sea
because they form a complete basis together. Now the question
arises how many levels in the Dirac sea one has to take into
account. In the calculations in Figs.~\ref{fig:sconvd.o16} and
\ref{fig:sconvd.pb208}, we have used $n_\text{max}^- =
n_\text{max}^+$ with $n_\text{max}^+$ determined by
$E_\text{cut}$. In Table~\ref{tab:convnn}, the dependence of the
average single particle energy, the rms radius $\langle
r^2\rangle^{1/2}$ and $\langle r^4\rangle^{1/4}$ on
$N_\text{max}^- = 2n_\text{max}^- + l$
--- a cutoff on the principal quantum number of levels in the
Dirac sea
--- are given for $^{16}$O and $^{208}$Pb. From
Table~\ref{tab:convnn}, we find a merit of solving the Dirac
equation in the DWS basis: the number of negative states included
in the basis could be much smaller that that of the positive
states. Let take $^{16}$O as an example, $R_\text{max} =
4r_0A^{1/3}$ and $E_\text{cut}$ = 300 MeV for positive states
correspond to $N_\text{max}^+ \sim$ 28. But for negative states,
$N_\text{max}^-$ = 10 gives very accurate results, e.g., the
discrepancy of $E^\text{sp}/A$ from the standard is smaller than
0.1 keV. This will significantly simplify the deformed problem by
decreasing the matrix dimension compared to solve the Dirac
equation in the SWS basis.

The above investigations are somehow academic. In practical
applications, it is not necessary to go to the accuracy around keV
in the single particle energy or 10$^{-4}$ fm in the radius. So in
the following calculations, we will use $R_\text{max}$ = 20 fm,
$\Delta r$ = 0.1 fm and $E_\text{cut}$ = 60$\sim$80 MeV for heavy
and light nuclei which give reasonable accuracies both for the
binding energy and the radius. This set of cutoff's corresponds
approximately $N_\text{max} = 2n_\text{max} + l \sim 25$ where $l$
is the orbital angular momentum of relevant state.

\section{\label{sec:results}Results and discussions}

In this section we present results of SRHWS. Since our main aim is
to show the virtues of SRHWS compared to SRHHO and SRHR, we do not
include pairing correlations and restrict our study to doubly
magic or magic nuclei only. If not specified, the parameter set
NLSH is used for the Lagrangian, $R_\text{max}$ = 20 fm and
$\Delta r$ = 0.1 fm throughout this section. Other parameter sets
for the Lagrangian do not change the conclusion here. In SRHDWS,
the number of normal levels in the Fermi sea and that of negative
ones in the Dirac sea are the same for convenience, i.e.,
$n_\text{max}^+ = n_\text{max}^-$. For SRHHO, $\hbar\omega_0 = 41
A^{-1/3}$ has been used and cutoff's of the expansion for fermions
and bosons are the same, i.e., $N_\text{F} = N_\text{B} \equiv
N_\text{max}$.

\subsection{\label{subsec:bulk}Bulk properties of stable nuclei
            from different SRH theories}

In Table~\ref{tab:bulk}, the binding energy per nucleon ($E/A$)
and neutron, proton and charge radii ($r_\text{n}$, $r_\text{p}$
and $r_\text{c}$) of some typical spherical nuclei are presented
which are calculated from the present available codes, including
SRHR, SRHSWS, SRHDWS, SRHHO. Available
data~\cite{Audi95,DeVries87} are also included for comparison. We
use approximately the same $N_\text{max}$ in the SRHHO as that in
the SRHWS which is determined by $E_\text{cut}$.

Generally speaking, for each studied nucleus, the four approaches
give almost the same results with an accuracy within 0.1\% with
few exceptions where the differences are still less than 0.3\%.
They are in excellent agreement with available data.

\begin{table}
\caption{\label{tab:bulk} The binding energy per nucleon and
neutron, proton and charge radii of some typical spherical nuclei.
The parameter set NLSH is used for the Lagrangian. $R_\text{max}$
= 20 fm and $\Delta r$ = 0.1 fm for SRHR and SRHWS. $\hbar\omega_0
= 41 A^{-1/3}$ for SRHHO. Numbers in brackets in the second column
give $E_\text{cut}$ for SRHWS and $N_\text{max}$ for SRHHO. Data
for $E/A$ and $r_\text{c}$ are taken from Ref.~\cite{Audi95} and
Ref.~\cite{DeVries87}, respectively. Energy is in MeV and radius
in fm.}
\begin{ruledtabular}
\begin{tabular}{llcccc}
 Nucleus    &        & $E/A$    & $r_\text{n}$
 & $r_\text{p}$ & $r_\text{c}$ \\
\hline
 $^{16}$O   & SRHR   & $-8.022$ & 2.551 & 2.578 & 2.699 \\
     & SRHSWS (80)   & $-8.022$ & 2.554 & 2.581 & 2.702 \\
     & SRHDWS (80)   & $-8.014$ & 2.553 & 2.580 & 2.701 \\
     &  SRHHO (25)   & $-8.034$ & 2.551 & 2.577 & 2.699 \\
     & Experiment    & $-7.976$ & & & 2.693 \\
\hline
 $^{40}$Ca  & SRHR   & $-8.500$ & 3.311 & 3.359 & 3.452 \\
     & SRHSWS (80)   & $-8.499$ & 3.310 & 3.358 & 3.452 \\
     & SRHDWS (80)   & $-8.497$ & 3.312 & 3.359 & 3.453 \\
     &  SRHHO (25)   & $-8.514$ & 3.310 & 3.358 & 3.452 \\
     & Experiment    & $-8.551$ & & & 3.478 \\
\hline
 $^{48}$Ca  & SRHR   & $-8.644$ & 3.586 & 3.369 & 3.463 \\
     & SRHSWS (80)   & $-8.646$ & 3.583 & 3.368 & 3.461 \\
     & SRHDWS (80)   & $-8.639$ & 3.586 & 3.371 & 3.464 \\
     &  SRHHO (25)   & $-8.659$ & 3.584 & 3.368 & 3.462 \\
     & Experiment    & $-8.666$ & & & 3.479 \\
\hline
 $^{56}$Ni  & SRHR   & $-8.634$ & 3.582 & 3.630 & 3.717 \\
     & SRHSWS (80)   & $-8.640$ & 3.580 & 3.628 & 3.715 \\
     & SRHDWS (80)   & $-8.625$ & 3.585 & 2.633 & 3.720 \\
     &  SRHHO (25)   & $-8.650$ & 3.581 & 3.629 & 3.716 \\
     & Experiment    & $-8.345$ & & & \\
\hline
%\end{tabular}
%\end{ruledtabular}
%\end{table}
%\begin{table}
%\begin{ruledtabular}
%\begin{tabular}{llcccc}
 $^{90}$Zr  & SRHR   & $-8.677$ & 4.294 & 4.186 & 4.262 \\
     & SRHSWS (75)   & $-8.677$ & 4.295 & 4.187 & 4.263 \\
     & SRHDWS (75)   & $-8.672$ & 4.295 & 4.187 & 4.262 \\
     &  SRHHO (25)   & $-8.693$ & 4.293 & 4.185 & 4.261 \\
     & Experiment    & $-8.710$ &       &       & 4.270 \\
\hline
 $^{118}$Sn & SRHR   & $-8.466$ & 4.743 & 4.553 & 4.623 \\
     & SRHSWS (70)   & $-8.466$ & 4.743 & 4.554 & 4.624 \\
     & SRHDWS (70)   & $-8.460$ & 4.743 & 4.554 & 4.624 \\
     &  SRHHO (25)   & $-8.482$ & 4.741 & 4.552 & 4.622 \\
     & Experiment    & $-8.517$ &       &       & 4.641 \\
\hline
 $^{132}$Sn & SRHR   & $-8.377$ & 4.964 & 4.636 & 4.704 \\
     & SRHSWS (70)   & $-8.377$ & 4.964 & 4.637 & 4.704 \\
     & SRHDWS (70)   & $-8.370$ & 4.964 & 4.637 & 4.706 \\
     &  SRHHO (25)   & $-8.393$ & 4.963 & 4.635 & 4.703 \\
     & Experiment    & $-8.355$ &       &       &       \\
\hline
 $^{208}$Pb & SRHR   & $-7.885$ & 5.713 & 5.447 & 5.505 \\
     & SRHSWS (60)   & $-7.886$ & 5.712 & 5.447 & 5.505 \\
     & SRHDWS (60)   & $-7.874$ & 5.712 & 5.448 & 5.506 \\
     &  SRHHO (25)   & $-7.900$ & 5.711 & 5.445 & 5.504 \\
     & Experiment    & $-7.868$ &       &       & 5.504 \\
\end{tabular}
\end{ruledtabular}
\end{table}

With the same parameters of spatial lattice $R_\text{max}$ and
$\Delta r$, SRHWS should reproduce results of SRHR when
$E_\text{cut}$ (or $N_\text{max}$) is large enough. This is
justified in Table~\ref{tab:bulk}. One find exactly coincident
results between SRHSWS and SRHR for most of the studied nuclei.
The remaining differences and those between SRHDWS and SRHR could
be diminished by increasing $E_\text{cut}$.

\begin{figure}
\includegraphics[width=8.5cm]{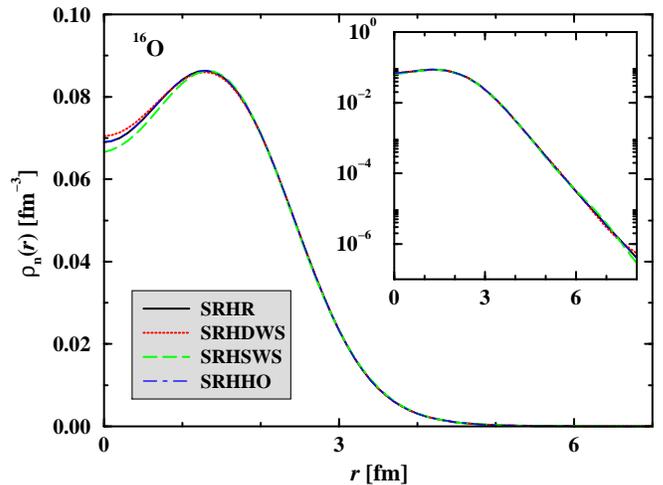}
\caption{\label{fig:o16rhon} Neutron density distributions for
$^{16}$O from different SRH approaches. The parameter set NLSH is
used for the Lagrangian. $R_\text{max}$ = 20 fm and $\Delta r$ =
0.1 fm for SRHR and SRHWS. $E_\text{cut} = 80$ MeV for SRHWS.
Correspondingly, $N_\text{max}$ = 25 for SRHHO. In SRHDWS, the
number of levels in the Dirac sea included in each block is the
same as that of normal levels which is determined by
$E_\text{cut}$. The inset presents logarithmic densities.}
\end{figure}

\begin{figure}
\includegraphics[width=8.5cm]{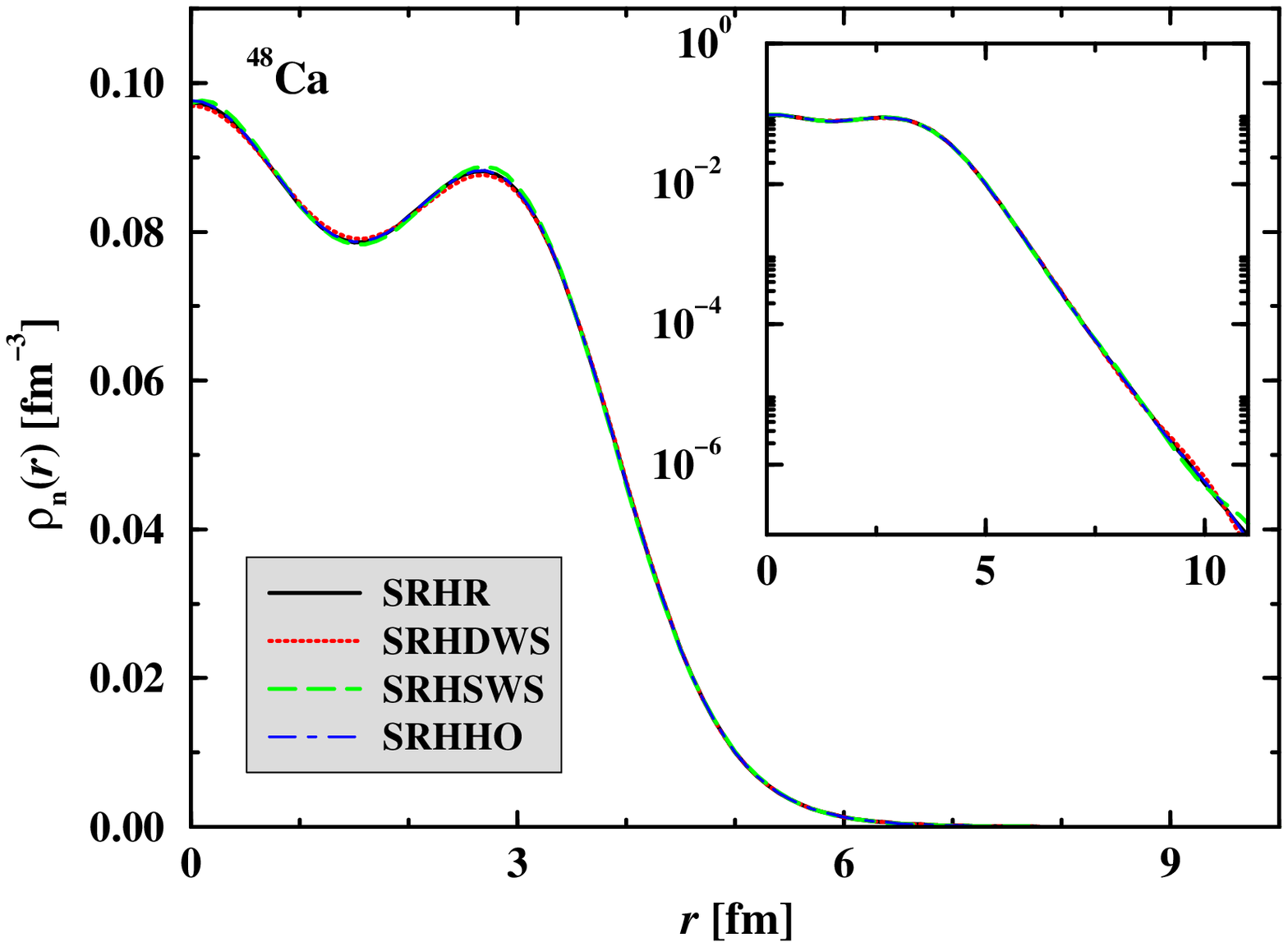}
\caption{\label{fig:ca48rhon} Neutron density distributions for
$^{48}$Ca from different SRH approaches. The parameter set NLSH is
used for the Lagrangian. $R_\text{max}$ = 20 fm and $\Delta r$ =
0.1 fm for SRHR and SRHWS. $E_\text{cut} = 80$ MeV for SRHWS.
Correspondingly, $N_\text{max}$ = 25 for SRHHO. In SRHDWS, the
number of levels in the Dirac sea included in each block is the
same as that of normal levels which is determined by
$E_\text{cut}$. The inset presents logarithmic densities.}
\end{figure}

\begin{figure}
\includegraphics[width=8.5cm]{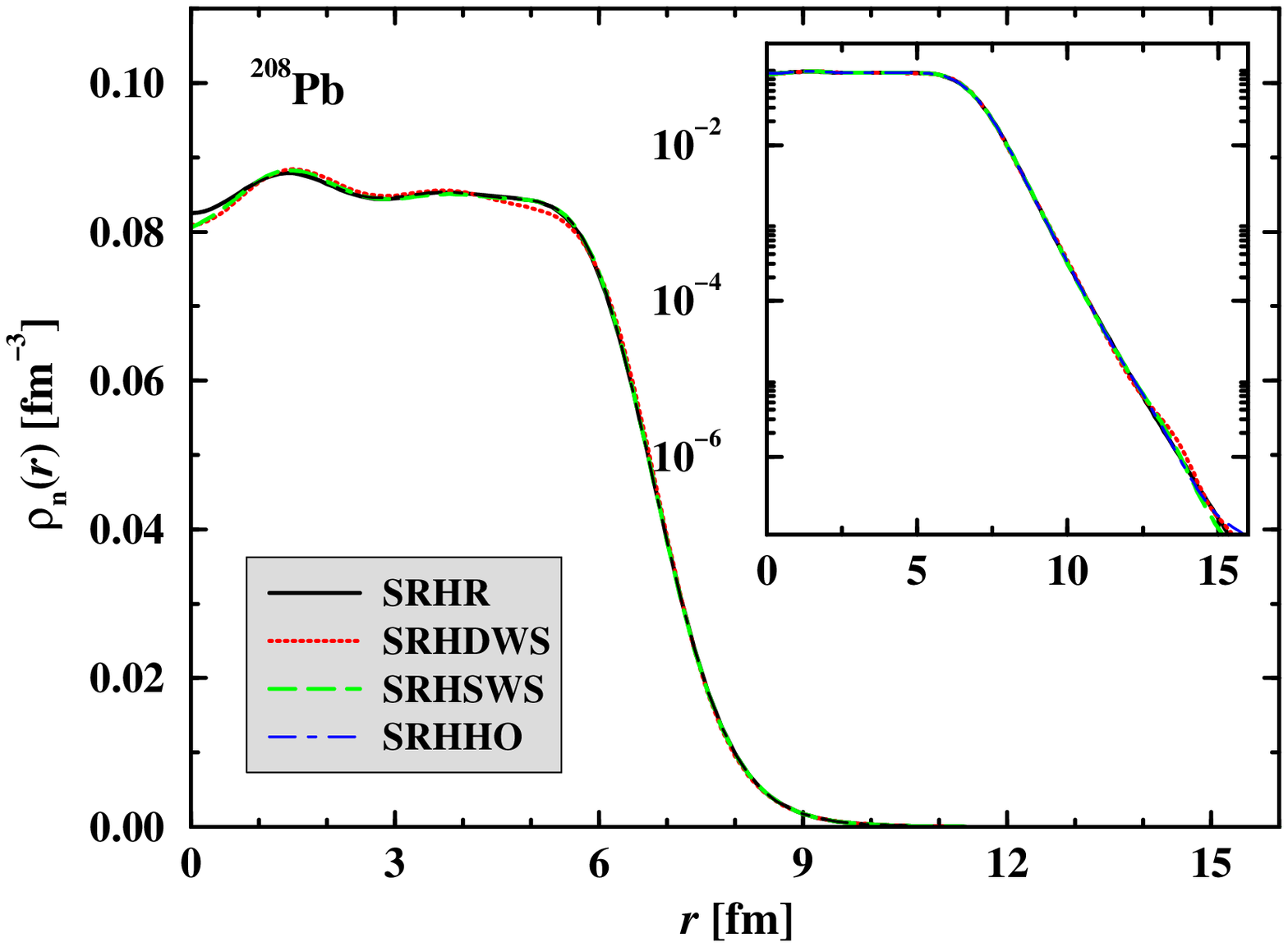}
\caption{\label{fig:pb208rhon} Neutron density distributions for
$^{208}$Pb from different SRH approaches. The parameter set NLSH
is used for the Lagrangian. $R_\text{max}$ = 20 fm and $\Delta r$
= 0.1 fm for SRHR and SRHWS. $E_\text{cut} = 60$ MeV for SRHWS.
Correspondingly, $N_\text{max}$ = 25 for SRHHO. In SRHDWS, the
number  of levels in the Dirac sea included in each block is the
same as that of normal levels which is determined by
$E_\text{cut}$. The inset presents logarithmic densities.}
\end{figure}

In Figures~\ref{fig:o16rhon}, \ref{fig:ca48rhon} and
\ref{fig:pb208rhon}, the neutron density distributions are
compared between SRHR, SRHSWS, SRHDWS, and SRHHO, in which
$^{16}$O, $^{48}$Ca and $^{208}$Pb are chosen as examples. The
calculation details are the same as Table~\ref{tab:bulk}. For
these stable nuclei, all these SRH methods are valid and all
calculations are in excellent agreement with each other from the
central to outer region of each nucleus. Tiny differences in the
central region do not affect the physical observables like the
binding energy or nuclear radius as is seen in
Table~\ref{tab:bulk}. Furthermore, these differences could also be
decreased by increasing $E_\text{cut}$ or $N_\text{max}$.

From the above discussions, it is clear that SRHWS is equivalent
to SRHR and SRHHO for stable nuclei. Thus we conclude that
Woods-Saxon basis provide another possibility to solve
(non-)relativistic mean field theory.

\subsection{\label{subsec:density}Neutron density distributions
            for $^{72}$Ca in different SRH theories}

As we already discussed in the introduction, one of the merits of
SRHR against SRHHO is its proper description of exotic nuclei. In
this subsection, we will demonstrate the equivalence between SRHWS
and SRHR when reasonably large $E_\text{cut}$ is applied in SRHWS.

In order to see the results for the unstable nuclei near the
neutron drip line, the neutron density distribution for $^{72}$Ca
is studied here. The nucleus $^{72}$Ca is predicted to be the last
bound calcium isotope~\cite{Im00,Hamamoto01,Zhang02,Meng02}. Since
it is not a doubly magic nucleus, there might be some uncertainty
in present results due to the lack of inclusion of pairing
correlations. However, as we stressed in the beginning of this
section, the main aim here is to show the virtue of SRHWS compared
to SRHHO, it is very unlikely that the pairing would change our
conclusion qualitatively.

\begin{table}
\caption{\label{tab:ca72.conv} Convergence study for $^{72}$Ca.
The parameter set NLSH is used for the Lagrangian. $\Delta r$ =
0.1 fm for SRHR and SRHWS. $E_\text{cut} = 75$ MeV for SRHWS.
Energy is in MeV and radius in fm.}
\begin{ruledtabular}
\begin{tabular}{ccccc}
                & $E/A$ & $r_\text{n}$ & $r_\text{p}$
 & $\lambda_\text{n}$ \\
\hline
 $R_\text{max}$ & \multicolumn{4}{c}{SRHR} \\
 20             & 6.482 & 4.656        & 3.639        & $-$0.191 \\
 25             & 6.483 & 4.723        & 3.639        & $-$0.221 \\
 30             & 6.484 & 4.773        & 3.639        & $-$0.228 \\
 35             & 6.484 & 4.807        & 3.639        & $-$0.229 \\
\hline
 $R_\text{max}$ & \multicolumn{4}{c}{SRHSWS} \\
 20             & 6.481 & 4.663        & 3.639        & $-$0.206 \\
 25             & 6.482 & 4.726        & 3.639        & $-$0.231 \\
 30             & 6.483 & 4.774        & 3.639        & $-$0.237 \\
 35             & 6.483 & 4.803        & 3.639        & $-$0.238 \\
\hline
 $R_\text{max}$ & \multicolumn{4}{c}{SRHDWS} \\
 20             & 6.474 & 4.662        & 3.641        & $-$0.163 \\
 25             & 6.475 & 4.733        & 3.641        & $-$0.197 \\
 30             & 6.475 & 4.789        & 3.640        & $-$0.205 \\
 35             & 6.475 & 4.828        & 3.640        & $-$0.206 \\
\hline
 $N_\text{max}$ & \multicolumn{4}{c}{SRHHO} \\
 25             & 6.489 & 4.577        & 3.639        & $-$0.054 \\
 31             & 6.492 & 4.605        & 3.639        & $-$0.128 \\
 37             & 6.494 & 4.628        & 3.639        & $-$0.166 \\
 43             & 6.494 & 4.649        & 3.639        & $-$0.189 \\
\end{tabular}
\end{ruledtabular}
\end{table}

For stable nuclei, it has been shown that $R_\text{max}\sim$ 20 fm
is large enough. For drip line nuclei, the dependence of the
results on $R_\text{max}$ for $^{72}$Ca is presented in
Table~\ref{tab:ca72.conv}.  For both SRHR and SRHWS, $\Delta r$ =
0.1 fm and $R_\text{max}$ = 20, 25, 30 and 35 fm have been adopted
respectively.  The energy cutoff $E_\text{cut}$ = 75 MeV is used
to SRHWS. In SRHR and SRHWS calculations, the neutron rms radius
$r_\text{n}$ and Fermi energy $\lambda_\text{n}$ of $^{72}$Ca
converge around $R_\text{max}$ = 35 fm while a independence of the
binding energy per nucleon $E/A$ and proton rms radius
$r_\text{p}$ on the box size can be seen. These sets of parameters
in SRHWS, $E_\text{cut}$ = 75 MeV and $R_\text{max}$ = 20, 25, 30
and 35 fm, correspond to cutoff's on principal quantum number
$N_\text{max}$ = 25, 31, 37 and 43 which are used in SRHHO
calculations in order to make fair comparisons between SRHWS and
SRHHO. Similar as those from SRHR and SRHWS, $E/A$ and
$r_\text{p}$ depend little on $N_\text{max}$ in SRHHO. However,
the neutron rms radius $r_\text{n}$ increases with $N_\text{max}$
steadily, which shows a much slower convergence. As it is based on
a complete basis, SRHHO can reach convergence of $r_\text{n}$ if
$N_\text{max}$ is large enough. From Table~\ref{tab:ca72.conv},
one finds that for the same $N_\text{max}$ (or equivalent
$R_\text{max}$), a difference of $\Delta r_\text{n} \approx$ 0.2
fm between SRHHO and SRHWS (SRHR) can be seen. From the slow
convergence of  $r_\text{n}$ with $N_\text{max}$ in SRHHO ($\Delta
N_\text{max}$ = 6 gives $\Delta r_\text{n} \approx$ 0.02 fm), we
can estimate the lower limit of the $N_\text{max}$ as
$N_\text{max} \approx$ 90 in order to give $r_\text{n}$ = 4.8 fm
in SRHR or SRHWS.

\begin{figure}
\includegraphics[width=8.5cm]{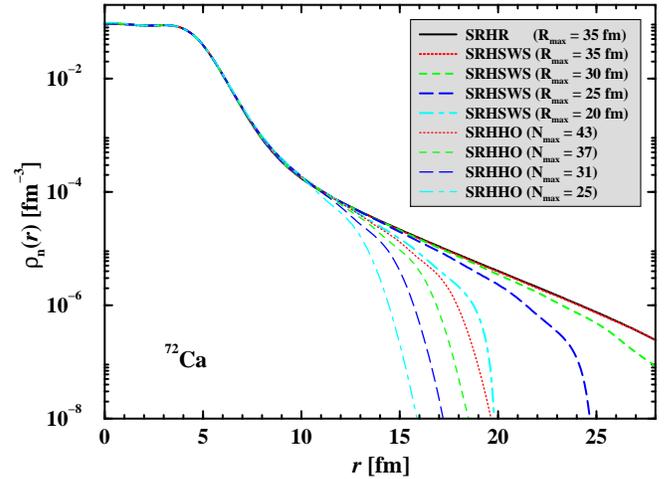}
\caption{\label{fig:ca72rhons} Comparison of density distributions
for $^{72}$Ca from SRHR, SRHSWS and SRHHO. The parameter set NLSH
is used for the Lagrangian. $\Delta r$ = 0.1 fm for SRHR and
SRHSWS. $E_\text{cut}$ = 75 MeV and $R_\text{max}$ = 20 (thick
dot-dashed curve), 25 (thick long-dashed curve), 30 (thick dashed
curve) and 35 fm (thick dotted curve) for SRHSWS. These sets of
cutoff's correspond to cutoff's in principal quantum number
$N_\text{max}$ = 25 (thin dot-dashed curve), 31 (thin long-dashed
curve), 37 (thin dashed curve) and 43 (thin dotted curve) which
are used in SRHHO calculations. The density distribution from SRHR
are almost identical with that from SRHSWS with the same box size.
For brevity, only $\rho_n(r)$ from SRHR with $R_\text{max}$ = 35
fm (thick solid line) is displayed here which covers the curve
corresponding to $\rho_n(r)$ from SRHSWS with $R_\text{max}$ = 35
fm (thick dotted curve).}
\end{figure}

\begin{figure}
\includegraphics[width=8.5cm]{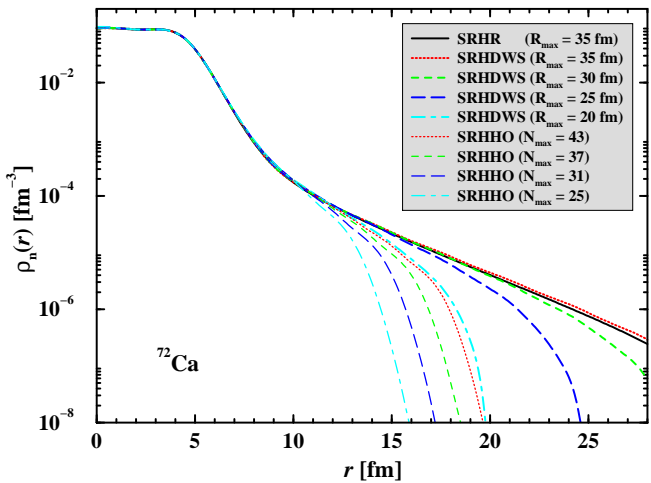}
\caption{\label{fig:ca72rhond} Comparison of density distributions
for $^{72}$Ca from SRHR, SRHDWS and SRHHO. The parameter set NLSH
is used for the Lagrangian. $\Delta r$ = 0.1 fm for SRHR and
SRHDWS. $E_\text{cut}$ = 75 MeV and $R_\text{max}$ = 20 (thick
dot-dashed curve), 25 (thick long-dashed curve), 30 (thick dashed
curve) and 35 fm (thick dotted curve) for SRHDWS. In SRHHO
calculations, $N_\text{max}$ = 25 (thin dot-dashed curve), 31
(thin long-dashed curve), 37 (thin dashed curve) and 43 (thin
dotted curve) are used. The density distribution from SRHR are
very close to that from SRHSWS with the same box size. For
brevity, only $\rho_n(r)$ from SRHR with $R_\text{max}$ = 35 fm
(thick solid line) is displayed here.}
\end{figure}

We compare the neutron density distribution of $^{72}$Ca from
different SRH approaches in Figs.~\ref{fig:ca72rhons} and
\ref{fig:ca72rhond}. With the same box size, the density
distribution from SRHR are almost identical with those from SRHWS,
which indicates the equivalence between SRHWS and SRHR. For
brevity, only $\rho_\text{n}(r)$ from SRHR with $R_\text{max}$ =
35 fm is displayed in Figs.~\ref{fig:ca72rhons} or
\ref{fig:ca72rhond} which covers the curve corresponding to
$\rho_n(r)$ from SRHSWS with $R_\text{max}$ = 35 fm in
Fig.~\ref{fig:ca72rhons}. On the other hand, $\rho_\text{n}(r)$
from SRHHO even with $N_\text{max}$ = 43 fails to reproduce the
result of SRHR due to the localization property of HO potential.
In addition, with the same $N_\text{max}$, the spatial extension
of $\rho_\text{n}(r)$ from SRHWS are always larger than that from
SRHHO. The variational tendency of the curve $\rho_\text{n}(r)
\sim r$ also explains different convergence behaviors of
$r_\text{n}$ in SRHWS and SRHHO as given in
Table~\ref{tab:ca72.conv}. With increasing $R_\text{max}$, the
$\rho_\text{n}(r)$ in SRHWS has the correct asymptotic behavior
while that in SRHHO decay too quickly.

This result is very encouraging and tell us that even the long
tail (or halo) behavior in neutron density distribution for nuclei
near the drip line can be described by SRHSWS as well as SRHR, if
pairing correlations is incorporated properly.

\subsection{\label{subsec:negative}Contribution from negative levels
            in the SRHDWS theory}

In the expansion of the nucleon wave function,
Eq.~(\ref{eq:DSexpansion}), one has to take into account not only
the levels in the Fermi sea but also those in the Dirac sea
because they form a complete basis together. We study the effects
of negative levels and give results of an example, $^{16}$O, in
Table~\ref{tab:o16.cn2}. Firstly, without negative levels
included, the nucleus is over bound and the nuclear size is
smaller as seen from Table~\ref{tab:o16.cn2}. Secondly, in
contrary to the case with negative levels included, the calculated
nuclear properties depend on the initial potentials very much if
no negative levels included.

It should be noted that the contribution from negative levels
depends on the initial Woods-Saxon potentials for generating the
DWS basis. So do the cutoff $N_\text{max}$ or $E_\text{cut}$ for
convergence. If the initial Woods-Saxon potential is exactly
identical to the converged potentials, the matrix in
Eq.~(\ref{eq:matrix}) is diagonal, negative states do not
contribute because of the no sea approximation. Positive states
can also be chosen as less as possible, e.g., $1s_{1/2}$,
$1p_{3/2}$ and $1p_{1/2}$ are enough for $^{16}$O. From the third
column corresponding to $V_0$ = 72 MeV in Table~\ref{tab:o16.cn2},
one finds that the initial nuclear potential for the Dirac
equation proposed in Ref.~\cite{Koepf91} is a good choice for
SRHDWS as the negative states only contribute $\sim$ 1.25 \% to
both $E/A$ and $r_\text{rms}$. If we change the initial
potentials, e.g., by changing $V_0$ by 25\%, much larger
contributions from negative states are found in
Table~\ref{tab:o16.cn2}.

\begin{table}
\caption{\label{tab:o16.cn2} Effects of negative levels on bulk
properties in SRHDWS for $^{16}$O. The parameter set for the
Lagrangian is NLSH, $R_\text{max} = 20$ fm, $\Delta r = 0.1$ fm,
and $N^+_\text{max} = 25$. For the initial Woods-Saxon like
potentials, parameters in Ref.~\cite{Koepf91} are used except for
$V_0$ which is specified in the table. The left value in each
entry gives the result without negative levels included and the
right one that with $N^-_\text{max}$ = 25. Energy is in MeV and
radius in fm.}
\begin{ruledtabular}
\begin{tabular}{lccc}
                & $V_0 = 54$ MeV  & $V_0 = 72$ MeV  & $V_0 = 90$ MeV  \\
\hline
 $E/A$          & 8.547 $|$ 8.013 & 8.117 $|$ 8.015 & 8.427 $|$ 8.012 \\
 $r_\text{rms}$ & 2.385 $|$ 2.568 & 2.531 $|$ 2.567 & 2.610 $|$ 2.567 \\
\end{tabular}
\end{ruledtabular}
\end{table}

\begin{table}
\caption{\label{tab:o16.cn1} Contribution of negative levels in
the Dirac sea to the single nucleon wave functions of $^{16}$O in
SRHDWS. For each single particle levels, $\sum |c^-_n|^2$ in the
expansion Eq.~(\ref{eq:DSexpansion}) is presented. The parameter
set for the Lagrangian is NLSH, $R_\text{max} = 20$ fm and $\Delta
r = 0.1$ fm and $N^+_\text{max} = N^-_\text{max} = 25$. For the
initial Woods-Saxon like potentials, the parameters in
Ref.~\cite{Koepf91} are used except for $V_0$ which is specified
in the table. The binding energy per nucleon and nuclear radius
for each calculation is presented in Table~\ref{tab:o16.cn2}.}
\begin{ruledtabular}
\begin{tabular}{lccc}
 Level       & $V_0 = 54$ MeV & $V_0 = 72$ MeV & $V_0 = 90$ MeV \\
\hline
 N$1s_{1/2}$& 9.15$\tm 10^{-5}$ & 3.48$\tm 10^{-5}$ & 3.44$\tm 10^{-4}$\\
 N$1p_{3/2}$& 5.32$\tm 10^{-5}$ & 6.05$\tm 10^{-5}$ & 3.69$\tm 10^{-4}$\\
 N$1p_{1/2}$& 4.01$\tm 10^{-4}$ & 1.06$\tm 10^{-5}$ & 8.80$\tm 10^{-4}$\\
 P$1s_{1/2}$& 1.92$\tm 10^{-4}$ & 4.92$\tm 10^{-5}$ & 4.89$\tm 10^{-5}$\\
 P$1p_{3/2}$& 1.17$\tm 10^{-4}$ & 3.27$\tm 10^{-5}$ & 8.07$\tm 10^{-5}$\\
 P$1p_{1/2}$& 8.09$\tm 10^{-4}$ & 2.08$\tm 10^{-4}$ & 2.38$\tm 10^{-5}$\\
\end{tabular}
\end{ruledtabular}
\end{table}

In order to know the contribution of negative levels in the Dirac
sea to the wave function, the value of $\sum_n |c_{n}^-|^2$ in the
expansion, Eq.~(\ref{eq:DSexpansion}) has been given in
Table~\ref{tab:o16.cn1} for occupied states of $^{16}$O. We note
that the nucleon wave function is normalized to one. It can be
seen that a small component from negative states in the wave
functions, about $10^{-4\sim -5}$, contributes to the physical
observables such as $E/A$ and $r_\text{rms}$ by the magnitude of
1\%$\sim$10\% as seen from Table~\ref{tab:o16.cn2}. Again we
notice that the initial Woods-Saxon potentials differ more from
the converged ones, the larger is the contribution from negative
levels.

\section{\label{sec:summary}Summary}

We have solved the spherical relativistic Hartree theory in the
Woods-Saxon basis (SRHWS). The Woods-Saxon basis is obtained by
solving either the Schr\"odinger equation (SRHSWS) or the Dirac
equation (SRHDWS). Formalism and numerical details for both cases
are presented. The WS basis in the SRHDWS theory could be very
smaller than that in the SRHSWS theory. This will largely
facilitate solving the deformed problem.

The results from SRHWS are compared with those from solving the
spherical relativistic Hartree theory in the harmonic oscillator
basis, SRHHO, and those in the coordinate space, SRHR. For stable
nuclei, all approaches give identical results for properties such
as total binding energies and the neutron, proton and charge rms
radii as well as neutron density distributions.

For neutron drip line nuclei, e.g. $^{72}$Ca, which has a very
small neutron Fermi energy $\lambda_\text{n} \sim 0.2$ MeV, both
SRHR and SRHWS easily approach convergence by increasing box size
while SRHHO does not. Furthermore, SRHWS can satisfactorily
reproduce the neutron density distribution from SRHR, but SRHHO
fails with similar cutoff's.

In SRHDWS calculations, the negative levels in the Dirac Sea must
be included in the basis in terms of which nucleon wave functions
are expanded. We studied in detail the effects and contributions
of negative states. Without negative levels included, the
calculated nuclear properties depend on the initial potentials
very much. A small component from negative states in the wave
functions, about $10^{-4\sim -5}$, contributes to the physical
observables such as $E/A$ and $r_\text{rms}$ by the magnitude of
1\%$\sim$10\%. When the initial potentials differ more from the
converged ones, the contribution from negative levels becomes more
important.

We conclude that the Woods-Saxon basis provides a reconciler
between the harmonic oscillator basis and the coordinate space
which may be used to describe exotic nuclei both properly and
efficiently.

The extension of spherical relativistic Hartree theory in the
Woods-Saxon basis to deformed cases with pairing included is in
progress.

\begin{acknowledgments}

S.-G. Z. would like to thank the Max-Planck-Institut f\"ur
Kernphysik for kind hospitality where part of this work is done.
J. M. would like to thank Physikdepartment, Technische
Universit\"at M\"unchen for kind hospitality. This work is partly
supported by the Major State Basic Research Development Program
Under Contract Number G2000077407 and the National Natural Science
Foundation of China under Grant No. 10025522, 10047001 and
19935030.

\end{acknowledgments}

\end{document}